\documentclass{article}

\usepackage{amsfonts}
\usepackage{epsfig}

\begin{document}

\title{\textsc{Intelligibility and First Passage Times In Complex Urban Networks}}

\vspace{1cm}

\author{Ph. Blanchard, D. Volchenkov
\vspace{0.5cm}\\
{\it  Bielefeld-Bonn Stochastic Centre},\\
{\it University Bielefeld, Postfach 100131,}\\
{\it D-33501, Bielefeld, Germany} \\
{\it Phone: +49 (0)521 / 106-2972 } \\
{\it Fax: +49 (0)521 / 106-6455 } \\
{\it E-Mail: VOLCHENK@Physik.Uni-Bielefeld.DE}}
\large

\date{\today}
\maketitle

\begin{abstract}

Topology of urban environments can be represented by means of graphs.
We explore the graph representations of several
 compact urban patterns by random walks.
The expected time of recurrence and the expected first passage
time to a node scales apparently linearly in all  urban patterns
we have studied In space syntax theory, a positive relation
between the local property of a node (qualified by connectivity or
by the recurrence time) and the global property of the node
(estimated in our approach by the first passage time to it) is
known as intelligibility. Our approach based on random walks
allows to extend the notion of intelligibility onto the entire
domain of complex networks and graph theory.

\end{abstract}


 \vspace{0.5cm}



\section{Spatial networks of urban environments}
\label{sec:Intro}
\noindent

More than half of all humans are now living in cities \cite{Ash:2008}.
Cities are responsible for a great deal of global energy consumption and of
greenhouse gas emissions.
The global challenges of sustainable development call for
a quantitative theory of urban organization.
There is a  well-established connection between the density of an urban environment,
 and the need to travel within it \cite{Wheeler:1998}.
 Good  quality of itineraries  is one of the necessary conditions for avoiding
  stagnation and collapse of a city.

Studies of urban transportation networks have a long history. In
his famous paper on the seven bridges of K\"{o}nigsberg published
in 1736, L. Euler had proven  the first theorem of graph theory
\cite{Biggs:1986}. In Euler's solution,  each urban landuse mass
is considered as a node of a planar graph, and the bridges
connecting them are the edges. Euler had found that a route
travelling along each edge in the planar graph representation of
the ancient K\"{o}nigsberg did not exists. In the  {\it primary}
graph representation of urban transport networks originated from
the work of Euler, the relationships between the different
components of urban environments are often measured along streets
and routes considered as edges, while the traffic end points and
street junctions are treated as nodes. In the last century,
primary city graphs have been used extensively in many studies
devoted to the improving of transportation routes, the
optimization of power grids, and the surveys of human mobility
patterns.

Another graph representation of urban transport networks
is based on the ideas of traffic engineering and queueing theory
 invented by A.K. Erlang
(see \cite{Brockmeyer:1948}).
It arises naturally when we are interested in how much time
 a pedestrian or a vehicle would spend while travelling
 through a particular place in a city.
In such a {\it secondary} graph representation,  any space
 of motion is considered as a service station of a queuing
 network characterized by some time of service, and the
relations between streets,  squares, and round-abouts are
traced through their  junctions.
Travellers arriving to a place  are either
moving through it immediately or queuing until the space becomes available.
Once the place is passed through, the traveller is routed to its
next station, which is chosen according to a probability distribution
among all other open spaces linked
to the given one in the urban environment.

In general, the secondary graph representations of urban environments are not planar.
Moreover, they are  essentially similar to those of
"dual information
representation" of a city map introduced by \cite{Rosvall:2005}
and to the "dual graphs"
extensively investigated within the concept of
{\it space syntax}, a theory developed in the late 1970s, that seeks
to reveal the mutual effects of complex spatial urban networks on
society and vice versa, \cite{Hillier:1984,Hillier:1999}.

In space syntax theory, built environments are treated as systems
of spaces of vision subjected to a configuration analysis.
Being
irrelevant to the physical distances, dual graphs representing the
urban environments are
removed from the physical space.
Spatial perception
shapes peoples understanding of how
a place is organized and eventually  determines the pattern of local
 movement, \cite{Hillier:1999}.
The aim
of the space syntax  study is to estimate the relative proximity
between different locations and to associate these distances to
 the densities of
human activity along the links connecting them, \cite{Hansen:1959,Wilson:1970,Batty:2004}.
The surprising accuracy of
predictions of human behaviour in cities based on the purely
topological analysis of different urban street layouts within
the space syntax approach attracts meticulous attention \cite{Penn:2001}.
Space syntax
proves its usefulness  for the planning and redevelopment of certain
city districts around the world, the designing of commercial centres,
 museums, railway stations, and airports where easy way-finding is a significant issue.

The decomposition of
urban spatial networks
into the complete sets
of intersecting open spaces
can be based on a number of different principles.
In  \cite{Jiang:2004},
while identifying a street over a plurality of routes
on a city map, the  named-street approach has been used, in
which two different arcs of the primary city network were
assigned to the same identification number (ID) provided they share the same
street name.
The main problem of the approach is that the
meaning of a street name could vary from one district or
quarter to another even within the same city.
For instance,
some streets in Manhattan do not meet the  continuity
principle rather playing the role of local geographical
coordinates.

In \cite{Porta:2006},
 an  intersection continuity principle (ICN) has been used: two
edges forming the  largest convex angle in a junction on the
city map are assigned the  highest continuity and therefore were
coupled together, acquiring the same street identification number (ID).
The main
problem with the ICN principle is that the streets crossing under
convex angles would artificially exchange their identifiers, which
 is not crucial for the study of the probability degree statistics
 \cite{Porta:2006}, but makes it difficult to interpret the results
 if the dynamical modularity of the city is detected \cite{Volchenkov:2007a}.
It is also important to mention that the number of street
IDs identified within the ICN principle usually exceeds substantially
the actual number of street names in a city.
In \cite{Cardillo:2006,Scellato:2006,Crucitti:2006},
the probability degree statistics and some centrality measures
in different
world cities have been investigated for a number of
one square mile representative samples.
 However, the decision on which a
square mile would provide an adequate representation of a
city is always questionable.

In \cite{Figueiredo:2005}, the approach of
\cite{Cardillo:2006,Scellato:2006,Crucitti:2006} has been improved
in the sense that
two intersecting lines of vision were aggregated into the same node
 of the dual graph representation
if and only if the angle
between the linear continuation of the first line and the second
line was less than or equal to a predefined threshold.
If more
than one continuation was available, the line forming the
smaller angle was chosen.

In our paper, we take a "named-streets"-oriented point of view
on the decomposition of
urban spatial networks
into the complete sets
of intersecting open spaces
 following our previous works \cite{Volchenkov:2007a,Volchenkov:2007b}.
 Being interested in the statistics of random walks defined on spatial
networks of urban patterns, we assign an individual
street ID code to each continuous segment of a street. The secondary
 graph is then constructed by
mapping all edges of the primary graph shared the same street ID into nodes
 and all intersections among each pair of edges of the primary graph
into the edges of the secondary graph connecting the corresponding nodes.

\section{The scope of the study and results}
\label{sec:Results}
\noindent

In the present paper, we explore the secondary graph representations
of different compact urban patterns
(the Venetian channel network,
the city of Paris, enclosed by the Peripheral Boulevard, and the almost regular
street grid in Manhattan)
by means or random walks.

In the forthcoming section (Sec.~\ref{sec:Connectivity}) we discuss the
connectivity statistics of secondary graphs representing urban environments.
In general, compact urban patterns have been developed under the deficits
of physical space
and therefore bear the multiple fingerprints of the physical
landscapes being scale dependent in general.
However, the large urban patterns which have not been
spatially restricted
during their development could constitute a
highly heterogeneous scalable
spatial networks as we demonstrate for the spatial network of the
city of Paris bounded by the Peripheral Boulevard.

In the Sec.~\ref{sec:Random_walks}, we explain how
random walks
can be used in order to
explore
complex networks.

 In
Subsec.~\ref{subsec:why_random_walks}, we demonstrate that the
transition operator of a random walk appears naturally as the
representation of the group of graph automorphisms in a class of
stochastic matrices.
Random walks provide us with an effective
tool for the detailed structural analysis of
connected undirected
graphs exposing their symmetries.
It is well known that while being defined on an undirected
graph, random walks determine
 a unique stationary probability distribution for every node \cite{Lovasz:1993}.
In Sec.~\ref{subsec:times}, we show that each node of a connected
undirected graph can be characterized with respect to random walks
by the expected recurrence time and the expected first passage
time. The expected recurrence time is simply the inverse of the
stationary probability distribution of random walks for the given
node and is therefore the local property of the node. The expected
first passage time figures out a global relation between the node
and other nodes of the given graph accounting for all possible
random paths toward the node accordingly to their respective
probabilities.

In Sec.~\ref{subsec:Embedding}, we show that for any
 undirected graph,
it is possible to define
a linear self-adjoint operator and
then use its nice spectral
properties  in order to extract
the information about the graph structure.
In particular, we
demonstrate that the complete set of orthonormal eigenvectors of
the symmetric transition operator can be used in order to introduce
the structure of Euclidean space on the graph.

In
Sec.~\ref{subsec:FPT_Euclidean}, we show that any node of the
graph can be represented as a vector in the $(N-1)$-dimensional
vector space and that Euclidean distances and angles between nodes
have  clear probabilistic interpretations.
 In particular, the
square of the norm of the vector representing a node
of the graph  equals to the expected first passage time
to it by random walkers.
 We can conclude that random walks embed
connected undirected graphs into
$(N-1)$-dimensional Euclidean
space.

The main result of our paper is explained in Sec.~\ref{sec:Intelligibility},
where we have shown that
the expected recurrence time scale apparently linear with the expected first passage times
in compact urban environments we have studied.

A similar strong positive relation between the local property of a place
 and its global properties with respect to other places in the dual graph
 of a city was known for a long time  in the framework of spaces
 syntax theory \cite{Hillier:1999}. Our  approach based on
 investigation of complex networks by means of random walks
 allows us to extend the notion of intelligibility far beyond
space syntax  onto the entire domains of complex networks and
the general theory of graphs.

\section{Connectivity statistics in secondary graphs representing urban environments}
\label{sec:Connectivity}
\noindent

The degree of a node representing a place
in the secondary graph representation of an urban environment
is the number of locations directly adjacent to the given one in the city.
 In space syntax theory,
the degree of a node (i.e. connectivity) is considered as
a local characteristic quantifying how well the space is connected
to others in the urban pattern,
\cite{Klarqvist:1993}.

 The probability degree distribution,
\begin{equation}
\label{degdistr01}
P(k)\,=\,\Pr\left[\,i\in G|\deg(i)\,=\,k\,\right],
\end{equation}
suggests
that a node selected uniformly at random has a certain degree $k$ with
 the probability $P(k)$.
The probability
degree distribution
is an important  concept characterizing the topology of complex networks.
It originates
from  the early studies of random graphs by Erd\"{o}s and R\'{e}nyi,
 \cite{Erdos:1959}{\it et al}
 as a common way to classify large graphs into categories such as
  {\it random graphs}  \cite{Erdos:1959}{\it et al} and scale-free
  networks \cite{Barabasi:2004}.

It has been reported earlier by \cite{Jiang:2004,Volchenkov:2007a} that
   the secondary graphs representing the urban environments
under the street-name identification principle
 exhibit the small-world property \cite{Newman:2001},
but the scale-free probability degree distributions
 pertinent to the scale-free graphs can hardly be recognized.
 In general,
compact city patterns do not provide us
with sufficient data to conclude on the universality of degree statistics.

\begin{figure}[ht]
 \noindent
\begin{center}
\epsfig{file=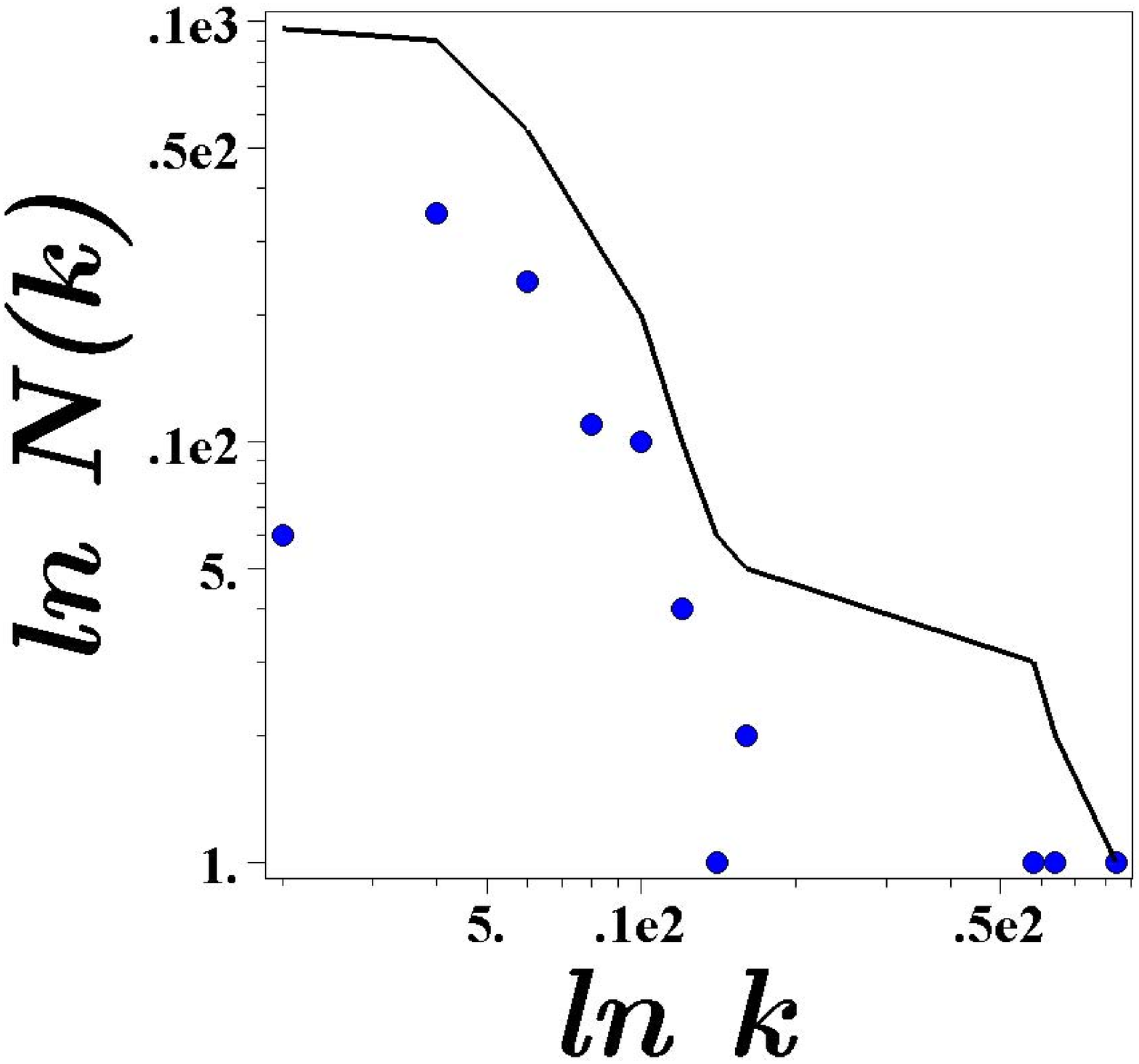, angle=0,width =7cm, height =7cm}
  \end{center}
\caption{\small  The log-log plot displays the
numbers of Venetian
 channels $N(k)$
versus the numbers $k$ of their
junctions.
The solid line indicates the cumulative
probability degree distribution, $N\mathfrak{P}(k).$ }
\label{Fig1_09}
\end{figure}

 To give an example, we display in Fig.~\ref{Fig1_09} the log-log plot of the numbers of
Venetian channels vs. the numbers of their junctions $k$.
The solid line indicates
 the cumulative empirical probability degree distribution,
 $N\mathfrak{P}(k)\,=\,\sum_{k'=k}^{N}\,N(k')/N,$
where $N=96$ is the total number of Venetian channels
(including those in the Giudecca island),
and $N(k')$ is the number of channels crossing precisely
$k'$ other channels.
It is remarkable
 that the empirical probability degree distributions
observed
for the secondary graphs
 are usually broad indicating that
the itineraries can cross different numbers of other routes.
Nevertheless,
the distributions usually
have a clearly recognizable maximum
corresponding to the most probable
number of junctions an average transport route has in the city.
The distributions usually have a
long right tail that decays faster then any
power law due to just a few
routes that cross many more others than in average, \cite{Volchenkov:2007a}.
This conclusion has been recently supported by
\cite{Figueiredo:2007} where it has been suggested that in general the
probability degree distributions of secondary graphs
  are scale-dependent.

\begin{figure}[ht]
 \noindent
\begin{center}
\epsfig{file=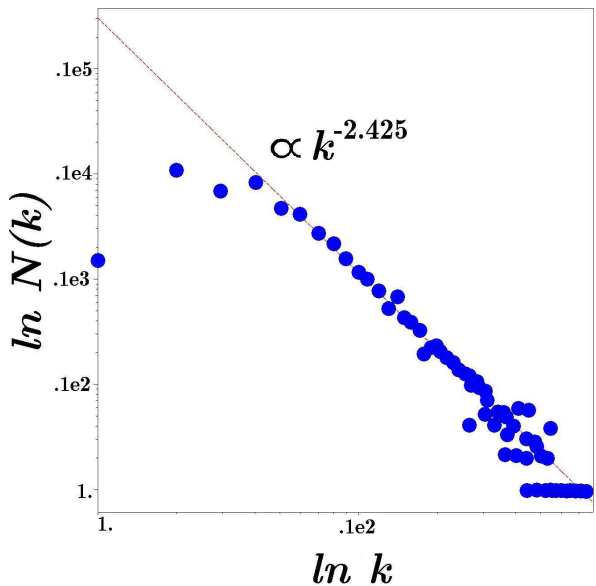, angle=0,width =8cm, height =8cm}
  \end{center}
\caption{\small  The log-log plot represents the empirical
degree statistics for the secondary graph of Paris
(5131 interconnected open spaces enclosed by the Peripheral Boulevard).
 $N(k)$ is the number of locations adjacent to exactly  $k$
other locations on the map of Paris.
The slope of the regression line equals $2.425$. }
\label{Fig2}
\end{figure}

It is important to note
that in the relatively large secondary graphs
which may contain many thousands of nodes a power law tail
can be observed in the probability degree distributions.
In Fig.~\ref{Fig2},
we have sketched the log-log plot of the numbers of open spaces
in the secondary graph of Paris (consisting of 5131
 interconnected open spaces
enclosed by the Peripheral Boulevard) versus
the numbers of their junctions with others.
The spatial network of Paris forms
 a highly heterogeneous apparently scalable graph.
In urban studies,
scaling and universality have been found  with a remarkable regularity.
The evolution
of social and economic life in cities increases with the population size:
wages, income, growth domestic product, bank deposits,
as well as the rate of innovations, measured by the number of new patents
and employment in creative sectors
scale super-linearly, over different years and nations,
with statistically consistent exponents \cite{Florida:2004,Bettencourt:2007}.

 The probable reason
for the similarity is that highly complex, self-sustaining
structures, whether cells, organisms, or cities constitute of an
immense  number of units are organized in a form of self-similar
hierarchical branching networks, which grow with the size of the
organism \cite{Enquist:1998}. A generic social dynamic underlying
the scaling phenomena observed in cities implies that an increase
in productive social opportunities, both in number and quality,
leads to quantifiable changes in individual behaviours of
inhabitants
 integrating them into a complex dynamical network \cite{Macionis:1998}.

The famous  rank-size distribution
 of city sizes  over many countries is known as  Zipf's Law \cite{Zipf:1949}.
 If we calculate the natural logarithm of the city rank in some
 countries and of the city size (measured in terms of its population)
 and then plot the resulting data in a diagram,  we obtain a remarkable
 linear pattern with the slope of the line equals $-1$ (or $+1$, if
  cities have been ranked in the ascending  order), \cite{Soo:2002}.
The similar centrality-rank distributions
for the values of a space syntax measure
quantifying the centrality of nodes
in the secondary graphs of
compact urban patterns have been recently reported by us in \cite{Volchenkov:2007b}.

\section{Exploration of graphs by random walks}
\label{sec:Random_walks}
\noindent

A graph naturally arises
as the outcome of a categorization,
when we abstract any real world system
by eliminating all but one of its features
and by grouping things (or places)
sharing a common attribute
by classes or categories.
For instance,
the common attribute
of all open spaces
in a city
is that
we can move through them.
All open spaces
found in a city
are considered as physically identical,
so that we can regard them
as nodes of a secondary graph $G(V,E)$,
in which $V$ is the set of all such spaces,
and $E$ is the set of all interconnections between them.
For each graph $G(V,E)$,
there  exists a unique,
up to permutations of rows and columns,
adjacency matrix $\mathbf{A}$.
In the special case
of a finite simple graph
(an undirected graph with no self-loops),
the adjacency matrix is a $\{0,1\}$-matrix
such that $A_{ij}=1$ if $i\ne j$,
  $i\sim j$,
and $A_{ij}=0$ otherwise.
The degree of a node $i\in V$ is therefore given  by
\begin{equation}
\label{condition}
k_i\,=\,\sum_{i\sim j}\,A_{ij}.
\end{equation}
For weighted undirected graphs,
the adjacency matrix $\bf A$ is replaced by
a symmetric positive affinity matrix $\bf w$.

\subsection{Why random walks?}
\label{subsec:why_random_walks}
\noindent

The set of graph automorphisms $\mathrm{Aut}(G)$,
the
mappings of the graph to itself
which preserve all of its structure,
 is
specified by the symmetric group $\mathbb{S}_N$ that includes all
admissible permutations $\Pi\in \mathbb{S}_N$ taking the node
$i\in V$ to some other node $\Pi(i)\in V$. The representation of
$\mathbb{S}_N$ consists of all  $N\times N$ matrices ${\bf
\Pi}_{\Pi},$ such that
  $\left({\bf \Pi}_{\Pi}\right)_{i,\,\Pi(i)}=1$,
and
 $\left({\bf \Pi}_{\Pi}\right)_{i,j}=0$
if $j\ne \Pi(i)$.

A linear transformation of the adjacency matrix
\begin{equation}
\label{lin_fun}
Z\left({\bf A}\right)_{ij}\,=
\,\sum_{\,s,l=1}^N\, \mathcal{F}_{ijsl}\,A_{sl}, \quad \mathcal{F}_{ijsl}\,\in \,\mathbb{R}
\end{equation}
belongs to $\mathrm{Aut}(G)$ if
\begin{equation}
\label{permut_invar}
{\bf \Pi}_{\Pi}^\top\, Z\left({\bf A}\right)\,{\bf \Pi}_{\Pi}\,=\,
Z\left({\bf \Pi}_{\Pi}^\top\,{\bf A}\,{\bf \Pi}_{\Pi}\right),
\end{equation}
for any $\Pi\in \mathbb{S}_N$.
It is clear that
the relation (\ref{permut_invar}) is satisfied
if the entries
of the tensor $\mathcal{F}$ in (\ref{lin_fun})
meet the following symmetry property:
 \begin{equation}
\label{symmetry}
\mathcal{F}_{\Pi(i)\,\Pi(j)\,\Pi(s)\,\Pi(l)}\,=\,
\mathcal{F}_{ijsl},
\end{equation}
for any $\Pi\in \mathbb{S}_N$.
Since the action of $\mathbb{S}_N$
 preserves the conjugate classes of index
partition structures,
it follows
that any appropriate tensor $\mathcal{F}$
satisfying (\ref{symmetry})
can be expressed
as a linear combination
of the following tensors:
$ \left\{1,
\delta_{ij},\delta_{is},\delta_{il},\delta_{js},\delta_{jl},\delta_{sl},
\delta_{ij}\delta_{js},\delta_{js}\delta_{sl},\right.\\
\left.
\delta_{sl}\delta_{li},
\delta_{li}\delta_{ij},
\delta_{ij}\delta_{sl},\delta_{is}\delta_{jl},\delta_{il}\delta_{js},
\delta_{ij}\delta_{il}\delta_{is} \right\}. $
By substituting
the above tensors into (\ref{lin_fun})
and taking account on
the symmetries,
 we obtain
that any
arbitrary linear permutation invariant
 function $Z\left({\bf A}\right)$
defined on a simple
undirected graph $G(V,E)$
must be of the following form,
\begin{equation}
\label{lin_fun2}
Z\left({\bf A}\right)_{ij}\,=\,a_1+\delta_{ij}\,\left(a_2+a_3k_j\right)+
a_4\,A{}_{ij},
\end{equation}
where  $k_j=\deg_G(j),$
and $a_{1,2,3,4}$ being arbitrary constants.

If we impose,
in addition,
that the
  linear function $Z$
preserves the  connectivity,
\begin{equation}
\label{conn_nodes}
k_i\,=\,\sum_{j\in V}\,Z\left({\bf A}\right)_{ij},
\end{equation}
it follows that  $a_1=a_2=0$ (since the contributions of $a_1N$
and $a_2$ are indeed incompatible with (\ref{conn_nodes})) and the
remaining constants should satisfy the relation $1-a_3=a_4$. By
introducing the new parameter $\beta\equiv a_4>0$, we can
reformulate (\ref{lin_fun2}) in the following form,
\begin{equation}
\label{lin_fun3}
Z\left({\bf A}\right)_{ij}\,=\, (1-\beta)\,\delta_{ij} k_j +
\beta\,A_{ij}.
\end{equation}
It is important to note
that
(\ref{conn_nodes})
can be interpreted as
a probability conservation relation,
\begin{equation}
\label{probab_nodes}
1\,=\,\frac 1{k_i}\sum_{j\in V}\,Z\left({\bf A}\right)_{ij}, \quad \forall i\in V,
\end{equation}
and therefore
the linear  function
$Z\left({\bf A}\right)$
can be interpreted as a stochastic process.

By substituting (\ref{lin_fun3}) into (\ref{probab_nodes}),
we obtain
\begin{equation}
\label{property}
\begin{array}{lcl}
1  & = & \sum_{j\in V}\, (1-\beta)\,\delta_{ij} +\beta\,\frac{A_{ij}}{k_i}\\
 & = & \sum_{j\in V}\, T^{(\beta)}_{ij},
\end{array}
\end{equation}
in which the operator $T^{(\beta)}_{ij}$
is nothing else but
 the generalized random
walk transition operator for $\beta\in[0,1]$.
The operator  $T^{(\beta)}_{ij}$
defines
"lazy" random walks
for which
 a random walker stays
in the initial vertex
with probability $1-\beta$,
while it moves
to another node
randomly chosen among the nearest neighbors
with
probability $\beta/k_i$.
In particular,
for  $\beta=1$,
 the operator
$T^{(\beta)}_{ij}$ describes the standard
random walks
extensively studied in classical surveys
\cite{Lovasz:1993},\cite{Aldous}.

\subsection{Recurrence and first passage times}
\label{subsec:times}
\noindent

Simple discrete time random walks
are the stochastic processes
where the position of a walker
at time $t$ depends
only on its position at time
$t-1$.
The attractiveness of random walks methods in
relies on the fact
that the distribution
of the current node of
any undirected  non-bipartite graph
after $t\gg 1$ steps
tends to a
well-defined stationary distribution $\pi$,
which is uniform
if the graph is regular.

Since the matrix $T^{(\beta)}_{ij}$ for any $\beta\in[0,1]$
is a real positive stochastic matrix, it follows from
 the
 Perron-Frobenius theorem \cite{Horn:1990} that
its maximal
 eigenvalue is simple and equals 1.
 The left eigenvector,
$
\pi\,T^{(\beta)}=\pi,
$
 associated with the maximal eigenvalue 1 is positive,
\begin{equation}
\label{stationary_pi}
\pi_i\,=\,\frac{k_i}{\sum_{i\in V}k_i},
\end{equation}
and satisfies
 the normalization condition $\sum_{i\in V}\pi_i=1$ independently of $\beta\in[0,1]$.
The left eigenvector $\pi$ is interpreted as the unique
 stationary distribution of this random walk.
If the graph $G$ is not bipartite,
any density function
$\sigma$ ($\sigma_i\geq 0$, $\sum_{i\in V} \sigma _i=1$)
asymptotically tends to the stationary
distribution $\pi$
under the actions of the transition operator  $T^{(\beta)}_{ij}$,
\begin{equation}
\label{limit}
\pi\,=\,\lim_{t\to\infty}\,\sigma\, \left(T^{(\beta)}\right)^{\,t}.
\end{equation}
Let us consider a random walk
\[\omega_{x_0}=\left\{x_0,x_1,\ldots x_n=i,\ldots x_{n+m}=i,\ldots \right\}\]
 starting from some node $x_0\in V$ chosen randomly among all nodes of the
 undirected graph $G$.
It is clear from (\ref{stationary_pi})
that for long enough random walks
the probability to find
a random walker in a certain node $i\in V$
equals $\pi_i$ that is
proportional to the degree of node $k_i$.
The expected recurrence time $\bar{m}_i$ to
 $i\in V$ is given by
$\bar{m}_i=\pi^{-1}_i$,
and therefore depends on
the local property of the node (its degree).

The first passage time $\bar{n}_i$ is
the
expected number of steps required for the random walker to reach the node
 $i$ for the first time  starting from a node randomly chosen among all
  nodes of the graph $G$.
This characteristic time is calculated as an average over all
random paths toward the node taken into account in accordance with
their respective probabilities. Being the global characteristic of
the node, $\bar{n}_i$
 estimates the level of accessibility to the node
from the rest of the graph.

Let us now calculate the first passage time to a node
by using spectral analysis of a self-adjoint transition operator.

Probably, Lagrange
was the first scientist
who investigated
a simple dynamical process
(diffusion) in order
to study the properties of a graph, \cite{Lag67}.
He calculated the spectrum
of the Laplace operator defined on a chain (a linear graph) of $N$ nodes in order to
 study the discretization
of the acoustic equations.
The idea of using
 the spectral properties of self-adjoint operators in order
to extract information about  graphs is standard in spectral graph theory
\cite{Chung:1997} and in theory of random walks on graphs \cite{Lovasz:1993},\cite{Aldous}.
In the  following calculations,
we take $\beta=1$
in the transition operator (\ref{property}).
This choice  allows us to compare our results
 directly
with those known from the classical surveys on random walks
\cite{Lovasz:1993},\cite{Aldous}.

\subsection{Euclidean embedding of graphs  by random walks}
\label{subsec:Embedding}
\noindent

The stationary distribution of random
walks $\pi$ defines a unique measure
on the set of nodes $V$
 with respect to which the
transition operator
((\ref{property}) for $\beta=1$)
is self-adjoint,
\begin{equation}
\label{self_adj}
\widehat{T}\,=\,\frac 12\left( \pi^{1/2}T\pi^{-1/2}+\pi^{-1/2}T^{\top}\pi^{1/2}\right),
\end{equation}
where
$T^\top$ is the adjoint operator,
and $\pi$ is defined as
the diagonal matrix
$\mathrm{diag}\left(\pi_1,\ldots,\pi_N\right)$.
In particular,
\begin{equation}
\label{T}
\widehat{T}_{ij}\, =\, \frac 1{\sqrt{k_ik_j}},\quad  \mathrm{iff}\,\, i\sim j,
\end{equation}
when $G$ is a simple undirected unweighted graph. The  ordered set
of real
 eigenvectors $\left\{\psi_i\right\}_{i=1}^N$ of the symmetric transition operator
 $\widehat{T}$
forms an orthonormal basis
in Hilbert space $\mathcal{H}(V)$.
The components of the first eigenvector
    $\psi_1$ belonging to the largest eigenvalue $\mu_1=1$,
\begin{equation}
\label{psi_01}
\psi_1
\,\widehat{ T}\, =\,
\psi_1,
\quad \psi_{1,i}^2\,=\,\pi_i,
\end{equation}
describes the connectivity of nodes.
The Euclidean norm
in the orthogonal complement of $\psi_1$,
$$\sum_{s=2}^N\psi_{s,i}^2\,=\,1-\pi_i$$,
gives the probability
that a random walker is not in $i$.
The  eigenvectors,
 $\left\{\,\psi_s\,\right\}_{s=2}^N$,
belonging to the ordered eigenvalues
  $1>\mu_2\geq\ldots\mu_N\geq -1$
 describe the connectedness
of the entire graph $G$.
The orthonormal system
of functions $\psi_s$ is useful for
 decomposing normalized functions
defined on $V$.

The symmetric transition operator
 $\widehat{T}$
projects
any density $\sigma\in \mathcal{H}(V)$ on
 the eigenvector $\psi_1$
related to the
  stationary distribution $\pi$,
$
\sigma\widehat{T}
=\psi_1 + \sigma_\bot\widehat{T}$, $ \sigma_\bot=\sigma-\psi_1,
$
in which $\sigma_{\bot}$ is the vector belonging to the orthogonal complement of
$\psi_1$ characterizing the transient
process toward the stationary distribution $\pi$ induced by $\sigma$.

Given two different densities $\sigma,\rho\in\mathcal{H}$, it is
clear that with respect to random walks they differ only on their
transient parts, but not on the final stationary state $\pi$.
Therefore, we can compare any two densities defined on $V$ by
means of random walks. Since all components $\psi_{1,i}>0$, it is
convenient to rescale the densities  by dividing their components
by  $\psi_{1,i}$,
\begin{equation}
\label{rescaling}
\widetilde{\sigma_i}\, =\,\frac{\sigma_i}{\psi_{1,i}}
\,=\, \frac{\sigma_i}{\sqrt{\pi_i}},\quad
\widetilde{\rho_i}\, =\,\frac{\rho_i}{\psi_{1,i}}
\,=\, \frac{\rho_i}{\sqrt{\pi_i}},
\end{equation}
such that for example $\widetilde{\sigma}\widehat{T}=1+\widetilde{\sigma}_\bot\widehat{T}$.
Then we define the squared Euclidean distance between any two densities
with respect to random walks by the sum over all times $t\geq 0$,
\begin{equation}
\label{distance}
\left\|\,\sigma-\rho\,\right\|^2_T\, =
\, \sum_{t\,\geq\, 0}\, \left\langle\, \widetilde{\sigma}_\bot -\widetilde{\rho}_\bot\,\left|\,\widehat{T}^t\,
\right|\, \widetilde{\sigma}_\bot -\widetilde{\rho}_\bot\,\right\rangle,
\end{equation}
where we have used  Dirac's bra-ket notations especially
convenient for working with inner products and
rank-one
operators in Hilbert space.
In order to perform the summation over time in (\ref{distance}), it is
convenient to
use the spectral
representation of $\widehat{T}$,
\begin{equation}
\label{spectral_dist}
\begin{array}{l}
\left\|\sigma-\rho\right\|^2_T
 \\
=
\sum_{t\,\geq 0} \sum_{s=2}^N\, \mu^t_s \left\langle
\widetilde{\sigma}^\bot -\widetilde{\rho}^\bot|\psi_s\right\rangle\!\left\langle \psi_s
| \widetilde{\sigma}^\bot -\widetilde{\rho}^\bot\right\rangle
\\
=
\sum_{s=2}^N\,\frac{\left\langle\, \widetilde{\sigma}_\bot -\widetilde{\rho}_\bot\,|
\, \psi_s\,\right\rangle\!\left\langle\, \psi_s\,
| \,\widetilde{\sigma}_\bot -\widetilde{\rho}_\bot\,\right\rangle}{\,1\,-\,\mu_s\,}.
\end{array}
\end{equation}
We conclude the description of the $(N-1)$-dimensional Euclidean
space structure induced by
  random walks defined on $G$ by mentioning that
every density  $\sigma\,\in\, \mathcal{H}(V)$ can be characterised
by the $(N-1)$-dimensional vector with the norm defined by
\begin{equation}
\label{sqaured_norm}
\left\|\, \sigma\,\right\|^2_T\,=\,
\,\sum_{s=2}^N \,\frac{\left\langle\,  \widetilde{\sigma}_\bot\,|\,\psi_s\,\right\rangle\!
\left\langle\,\psi_s\,|\, \widetilde{\sigma}_\bot\, \right\rangle}{\,1\,-\,\mu_s\,}.
\end{equation}
Moreover, given two densities $\sigma,\rho\,\in\, \mathcal{H}(V),$
we can introduce a scalar product
in the  $(N-1)$-dimensional Euclidean
space by
\begin{equation}
\label{inner-product}
\left(\,\sigma,\rho\,\right)_{T}
\,= \,  \sum_{s=2}^N
\,\frac{\,\left\langle\,  \widetilde{\sigma}_\bot\,|\,\psi_s\,\right\rangle\!
\left\langle\,\psi_s\,|\, \widetilde{\rho}_\bot \right\rangle}{\,1\,-\,\mu_s\,},
\end{equation}
so that
the
angle between  $\sigma,\rho\,\in\, \mathcal{H}(V)$
 can be calculated as
\begin{equation}
\label{angle}
\cos \,\angle \left(\rho,\sigma\right)=
\frac{\,\left(\,\sigma,\rho\,\right)_T\,}
{\left\|\,\sigma\,\right\|_T\,\left\|\,\rho\,\right\|_T}.
\end{equation}

\subsection{First passage time as the Euclidean norm of a node}
\label{subsec:FPT_Euclidean}
\noindent

Random walks embed connected undirected graphs into the Euclidean
space $\mathbb{R}^{N-1}$. This embedding  can be used directly in order to
calculate the first passage times to individual
nodes.

Indeed, let us consider the vector $\mathbf{e}_i=\{0,\ldots 1_i,\ldots 0\}$ that represents
the node $i\in V$ in the canonical basis as a density function.
In accordance to (\ref{sqaured_norm}), the vector $\mathbf{e}_i$
has the squared norm
of  $\mathbf{e}_i$ associated to random walks is
\begin{equation}
\label{norm_node}
\left\|\,\mathbf{e}_i\,\right\|_T^2\, =\,\frac 1{\pi_i}\,\sum_{s=2}^N\,
\frac{\,\psi^2_{s,i}\,}{\,1-\mu_s\,}.
\end{equation}
It is important to note that in the theory of random walks
\cite{Lovasz:1993} the r.h.s. of (\ref{norm_node}) is known
 as the spectral representation of the first passage time
$\bar{n}_i$
 to the node $i\in V$ from a node randomly chosen among all nodes
 of the graph accordingly to the stationary distribution $\pi$.
The first passage time, $\bar{n}_i=\left\|\mathbf{e}_i\right\|_T^2$,
 can be directly used in order to characterize
the level of accessibility of the node $i$.

The Euclidean distance between any two nodes
of the graph $G$ calculated in the $(N-1)-$dimensional
Euclidean space associated to random walks,
\begin{equation}
\label{commute}
K_{i,j}\,=\,\left\|\,\mathbf{e}_i-\mathbf{e}_j\,\right\|^2_T\,=\, \sum_{s=2}^N\,
\frac 1{1-\mu_s}\left(\frac{\psi_{s,i}}{\sqrt{\pi_i}}-\frac{\psi_{s,j}}{\sqrt{\pi_j}}\right)^2,
\end{equation}
also gets
a clear probabilistic interpretation
as the spectral representation of the commute time,
the expected number of steps required for a random
walker starting at $i\,\in\, V$ to visit $j\,\in\, V$ and then to
return back to $i$,  \cite{Lovasz:1993}.

The commute time can be represented as a sum,
$K_{i,j}=H_{i,j}+H_{j,i}$, in which
\begin{equation}
\label{hitting}
H_{i,j}\,=\,\left\|\,\mathbf{e}_i\,\right\|^2_T - \left(\,\mathbf{e}_i,\mathbf{e}_j\,\right)_T
\end{equation}
is the first hitting time which
quantifies the expected number of steps
a random walker starting from the node $i$ needs to reach
$j$ for the first time,  \cite{Lovasz:1993}.

The scalar product $\left(\mathbf{e}_i,\mathbf{e}_j\right)_T$ estimates the
expected overlap of random paths towards the nodes $i$ and $j$ starting from a
node randomly chosen in accordance with the stationary distribution of random walks $\pi$.
The normalized expected overlap of random paths given by the cosine of an angle calculated in
the $(N-1)-$dimensional Euclidean space associated to random walks
 has the structure of
Pearson's coefficient of linear correlations
 that reveals it's natural
statistical interpretation.
If the cosine of an angle (\ref{angle}) is close to 1
(zero angles),
it indicates that
the expected random paths toward both nodes are mostly identical.
The value of cosine is close to -1 if the walkers
share the same random paths but in the opposite direction.
Finally, the  correlation coefficient equals 0
if the expected random paths toward the nodes do not overlap.
It is important to mention that
 as usual the correlation between nodes
does not necessary imply a direct causal
relationship (an immediate connection)
between them.

\section{The first passage time and intelligibility of complex urban networks}
\label{sec:Intelligibility}
\noindent

It is intuitive that
the time of recurrence  to a node,
$\bar{m}_i$,
has to be positively related to the first passage time to it, $\bar{n}_i$:
the faster a random walker
 hits the node for the first time,
the more often he is expected to
visit it in future.
This intuition is supported by  (\ref{norm_node})
from
which it follows
 that $\bar{n}_i\propto\bar{m}_i$
provided the sum $\sum_{s=2}^N
\psi^2_{s,i}/\left(1-\mu_s\right)$
is  uniformly independent
of the connectivity for all nodes.
The possible relation between the local and global properties of nodes is
the most profound feature of a complex network.
It is interesting to note that
this nontrivial
 property of eigenvectors
seems to be true
for the secondary graphs representing
complex urban networks:
the first passage times to the nodes
scale apparently linearly with their connectivity.

\begin{figure}[ht]
 \noindent
\begin{center}
\epsfig{file=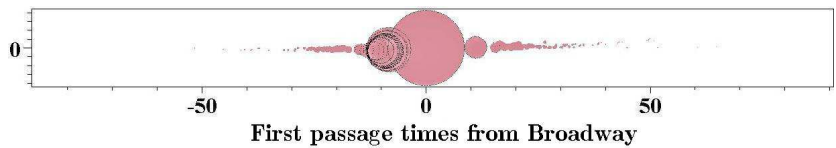, angle=0,width =11cm, height =2cm}
  \end{center}
\caption{\small  The 2-dimensional projection of
the Euclidean space of 355 locations
in Manhattan (New York) from Broadway set up by random walks.}
\label{Fig3}
\end{figure}

In Fig.~\ref{Fig3}, we have sketched the 2-dimensional projection
of the Euclidean space of 355 locations in Manhattan (New York)
set up by random walks. Nodes of the secondary graph are shown by
disks with radiuses taken proportional to the connectivity of the
places. Broadway, a wide avenue in Manhattan
 which also runs into the Bronx and Westchester County,
possesses the highest connectivity and located at the centre of the graph
shown in Fig.~\ref{Fig3}. Other places are located
at their Euclidean distances from Broadway
calculated accordingly (\ref{commute}), and (\ref{angle}) has been
used in order to compute angles
between Broadway and other places.

A part-whole relationship between local and global properties of the spaces of motion
is known in space syntax theory as an {\it intelligibility} property of
urban pattern \cite{Hillier:1984},\cite{Hillier:1999}. The adequate level of intelligibility
is proven to be a key determinant of the human behaviour
 in urban environments encouraging peoples way-finding abilities, \cite{Jiang:2004}.
In space syntax theory, the local property of an open space is
qualified by its connectivity, while its global property is
estimated by a special space syntax measure called 'integration'.
In the traditional space syntax analysis \cite{Hillier:1999}, the
integration of a place into urban environments is estimated by the
normalized sum of all graph theoretical distances toward the place
from all other places in the city. In \cite{Jiang:2004}, the
integration of a place has been estimated by means of the
centrality of the node in the dual graph representation of the
urban environment.

\begin{figure}[ht]
 \noindent
\begin{center}
\epsfig{file=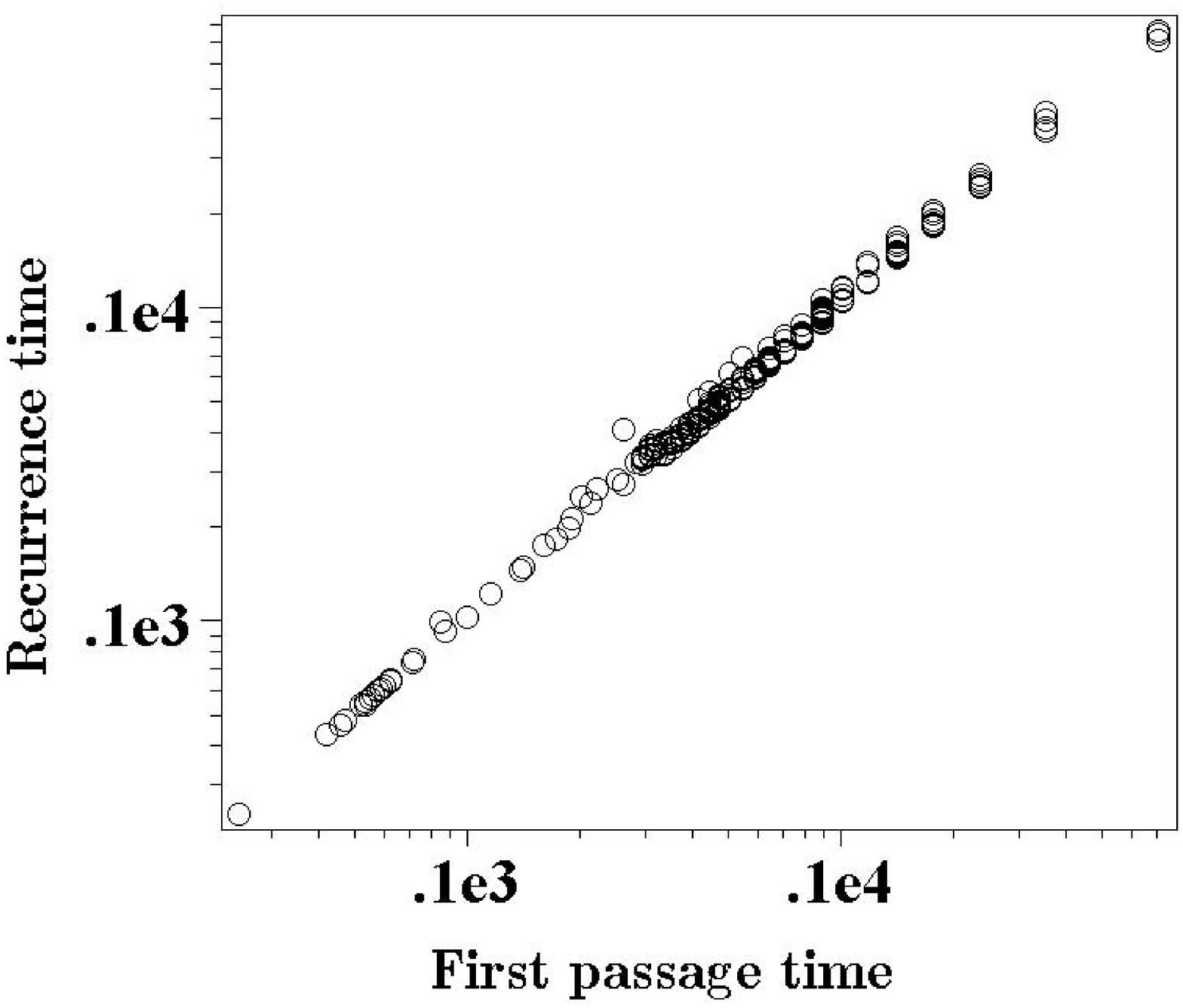, angle=0,width =7cm, height =7cm}
  \end{center}
\caption{\small  The times of recurrence to locations
in Manhattan
scale linearly with
the first passage times to them. }
\label{Fig4}
\end{figure}
The first passage time to a node which we use in the present paper in
order to quantify the relation
of the node with other nodes in the graph has an immediate connection
to neither the traditional space syntax integration measure discussed
 in \cite{Hillier:1984},\cite{Hillier:1999} nor the centrality measure
  investigated in \cite{Jiang:2004}.
However, the first passage time indicates
a strong positive relation between the local and global properties of
the spaces of motion in urban environments (see Fig~\ref{Fig4}) in a
 pretty same way as it has been demonstrated in the classical space
 syntax analysis.

The approach based on investigation of complex networks by means of random walks
 allows us to extended the notion of intelligibility far beyond the urban studies
where it has been originally invented onto the entire domains of complex networks and
the general theory of graphs.



\section{Acknowledgment}
\label{Acknowledgment}
\noindent

The work has been supported by the Volkswagen Foundation (Germany)
in the framework of the project "{\it Network formation rules, random
set graphs and generalized epidemic processes}" (Contract no Az.:
I/82 418).

\end{document}